\documentclass[sigconf]{acmart}

\usepackage{graphicx}
\usepackage{hyperref}
\usepackage{color}

\usepackage{subfigure}

\usepackage{algorithm}
\usepackage{algpseudocode}
\usepackage{lipsum}

\renewcommand\footnotetextcopyrightpermission[1]{} 
\begin{document}
\title{TDML - A Trustworthy Distributed Machine Learning Framework}

\author{Zhen Wang, Qin Wang, Guangsheng Yu, Shiping Chen}
\affiliation{
\textit{CSIRO Data61, Australia}
}

\begin{abstract}

Recent years have witnessed a surge in deep learning research, marked by the introduction of expansive generative models like OpenAI's SORA and GPT, Meta AI's LLAMA series, and Google's FLAN, BART, and Gemini models. However, the rapid advancement of large models (LM) has intensified the demand for computing resources, particularly GPUs, which are crucial for their parallel processing capabilities. This demand is exacerbated by limited GPU availability due to supply chain delays and monopolistic acquisition by major tech firms. Distributed Machine Learning (DML) methods, such as Federated Learning (FL), mitigate these challenges by partitioning data and models across multiple servers, though implementing optimizations like tensor and pipeline parallelism remains complex. Blockchain technology emerges as a promising solution, ensuring data integrity, scalability, and trust in distributed computing environments,  but still lacks guidance on building practical DML systems. In this paper, we propose a \textit{trustworthy distributed machine learning} (TDML) framework that leverages blockchain to coordinate remote trainers and validate workloads, achieving privacy, transparency, and efficient model training across public remote computing resources. Experimental validation demonstrates TDML's efficacy in overcoming performance limitations and malicious node detection, positioning it as a robust solution for scalable and secure distributed machine learning.

\end{abstract}


\maketitle    
\section{Introduction}





There has been a remarkable surge in deep learning research and its practical applications. Leading tech giants have unveiled their expansive generative models, with examples like OpenAI's SORA and GPT. Meta AI unveiled the LLAMA series, the world's first open-source large language model. Google introduced its language models, such as FLAN, BART, and Gemini. 

With the rapid advancement of large models (LM), computing resources have become a critical bottleneck in the AI domain. Graphics Processing Units (GPUs) are favored for AI computing due to their parallel infrastructure and ability to process data simultaneously, making them indispensable for machine learning tasks. However, the limited number of companies involved in GPU development and distribution creates significant delays in the manufacturing supply chain. Moreover, major tech companies in cloud computing exacerbate the shortage of computing resources for smaller organizations by prioritizing the acquisition of the majority of GPUs. For example, OpenAI and Microsoft plan to invest USD 100b in GPUs by 2027 to enhance their data center capabilities. Meta's Llama 3 models are trained on two clusters, each with 24,576 H100 GPUs, and Meta intends to acquire an additional 350,000 Nvidia H100 GPUs for over USD 10b. This unequal competitive landscape hampers the ability of AI startups to construct large deep learning models and compete on an even playing field.

Distributed Machine Learning (DML) integrates distributed computing resources to provide fast learning capabilities, particularly for tasks involving large-scale data or extensive model parameters. This method partitions the training data and the model, with parameter servers coordinating multiple clients to learn each partition as a subtask. Federated Learning (FL)~\cite{kairouz2021advances} exemplifies data parallelism in DML, coordinating the distributed training process using local data and aggregating a global model on a central server. Model parallelism training methods have been applied to solve many real problems that deal with large model-distributed training systems. The training network systems involve numerous connected computing and storage units. In order to reduce the model training complexity, various optimizations have been proposed to efficiently distribute the training across multiple devices, including tensor, data, and pipeline parallelism, as well as activation checkpointing~\cite{pippy2022, Rasley2020DeepSpeedSO, DBLP:journals/corr/abs-1909-08053, DBLP:journals/corr/abs-1910-02054}. However, most optimizations require changes to the model implementation manually, which overburdens model development and maintenance. For example, the checkpointing~\cite{DBLP:journals/corr/abs-2006-09616} needs to manually load the intermediate activation memories under a restricted memory situation. To support DeepSpeed model parallelism or pipeline parallelism, it requires integrating additional configurations from the DeepSpeed API, such as injection policies, into the model development process. Moreover, existing frameworks lack design for an open, flexible remote training solution and are only compatible with managed environments, resulting in expensive costs and limited resource availability. 

Blockchain technology~\cite{wood2014ethereum} operates by linking blocks together in a sequential manner to store transaction records, utilizing cryptographic algorithms to guarantee the immutability and authenticity of the blocks. It serves as a collaborative trust mechanism in distributed computing environments, emerging as a highly promising technology for security, scalability, and trust establishment~\cite{nguyen2023blockchain,wang2019artchain,nguyen2023bdsp}. A key feature of blockchain is its partition-block data structure, where data is distributed across partitions and each block records the identity document of the preceding one, ensuring the authenticity and immutability of data within each block. By leveraging blockchain, data reliability in either distributed cloud computing or model training~\cite{sai2024your} can be ensured while also providing protection against tampering~\cite{nguyen2023blockchain}. The integration of blockchain with smart contracts~\cite{li2022smart} enables users to perform authentic and traceable transactions without the need for a central intermediary.


Blockchain-based federated learning (BFL)~\cite{10.1145/3659099,10.1016/j.csi.2021.103561,qu2022blockchain,zhu2023blockchain} integrates the decentralized nature of blockchain with the distributed architecture of FL, mitigating the risk of a single point of failure in the FL system's aggregation server. BFL has been broadly studied in various domains, including mobile edge computing~\cite{nguyen2021federated}, internet of things~\cite{10.1016/j.cose.2021.102355,lu2019blockchain}, and distributed machine learning~\cite{article111,yu2023ironforge}. In particular, architectures like BlockFL~\cite{DBLP:journals/corr/abs-1808-03949} rely on a custom blockchain to exchange and validate local learning model updates. Similarly, Lu et al.~\cite{8843942} propose a system consisting of a dual module containing a permission Blockchain module and an FL module. 

\smallskip

Recent studies on blockchain-based distributed training systems have primarily focused on data parallelism methods, which are suitable only for standalone models with local edge data. The potential to leverage public distributed computing resources for large model training has not yet been explored. In this paper, we introduce a blockchain-based distributed framework designed to train expansive models using public remote computing resources. 

\smallskip
Our main \textit{\textbf{contributions}} are as follows:

\begin{itemize}

    \item We propose TDML (Sec.\ref{sec-model}), a new framework that utilizes blockchain technology to coordinate and verify the workloads of remote trainers in distributed ML to achieve privacy, transparency, and data traceability.
    
    \item To overcome the Byzantine attacks from malicious nodes, we also proposed malicious detection and incentive-driven mechanisms, respectively.

    \item We conduct comprehensive experiments (Sec.\ref{sec-exp}) showcasing the effectiveness of the proposed framework TDML against three baseline techniques using the ResNet50 model on the Cifar-10 dataset. These experiments aim to demonstrate that our framework can match the performance of single-node training, achieve comparable accuracy to traditional methods despite the challenges of federated learning, and enhance efficiency in terms of convergence speeds and training loss.
\end{itemize}

Experimental results show that our TDML framework surpasses the performance limitations of FedAvg and matches the baseline performance of single-node training. Additionally, the gradient-based malicious detection effectively identifies and isolates malicious trainers from benign nodes.

\section{Background and Related Work}\label{sec-rw}

\subsection{DNN Training Parallelism}
Large language models have demonstrated remarkable performance in a wide range of downstream tasks. In recent years, numerous large language models have been developed and released ~\cite{chiang2024chatbot, DBLP:journals/corr/abs-2110-14168}, demonstrating that larger models can generalize better across a wider range of tasks. However, as data volume and model complexity grow, it becomes increasingly difficult for a single machine to efficiently manage memory limitations. Distributed optimization and inference are essential to overcome large-scale DNN challenges. 

Two common distribution strategies include: i. distributing the training data across multiple computing nodes (\textit{data parallelism}); ii. dividing the large model across multiple computing nodes (\textit{model parallelism}).
Data parallelism boosts system throughput by replicating the same model across multiple computing nodes, with each node processing different chunks of data concurrently. This leads to higher throughput as multiple tasks are processed simultaneously, thereby reducing the overall time needed to complete the computations.  On the other hand, model parallelism is used when the model is too large to fit on a single computing node. This approach involves partitioning the model into segments across multiple nodes, with each node handling a portion of the model's computations. Consequently, the model's parameters are distributed, addressing the limited memory capacity issue.

\subsubsection{Data parallelism}

Beyond the traditional data parallelism of model mirror, an efficient memory management method, Zero Redundancy Optimizer (ZeRO)~\cite{DBLP:journals/corr/abs-1910-02054}, reduces the memory consumption of computing nodes by slicing the model states of model params, gradients and optimizer states into different computing nodes, instead of replicating all of them in each node. Scaling Fully Sharded Data Parallel (FSDP)~\cite{10.14778/3611540.3611569}, similar to ZeRO, distributes a model’s parameters, gradients, and optimizer states across computing workers. It enables parameters to be stored in CPU memory and only loads the parameters for the current computing layer. It further reduces memory usage. However, the data parallelism strategies of sharing parameters across computing nodes are not suitable for our case (training a large model across untrusted workers). Additionally, fully leveraging the benefits of ZeRO or FSDP requires significant manual tweaking during the training process to efficiently send and gather model parameters.

Federated learning is another intriguing application of data parallelism. It reduces data transmission costs by averaging locally trained models to create a global model. While this method enhances data privacy, it can lead to a suboptimal global model and is not well-suited for training large models.


\begin{figure}[!htb]
\centering
\subfigure[Data parallelism]{\label{fig:data_paral}
\resizebox{0.99\linewidth}{!}{
	\includegraphics[width=\textwidth]{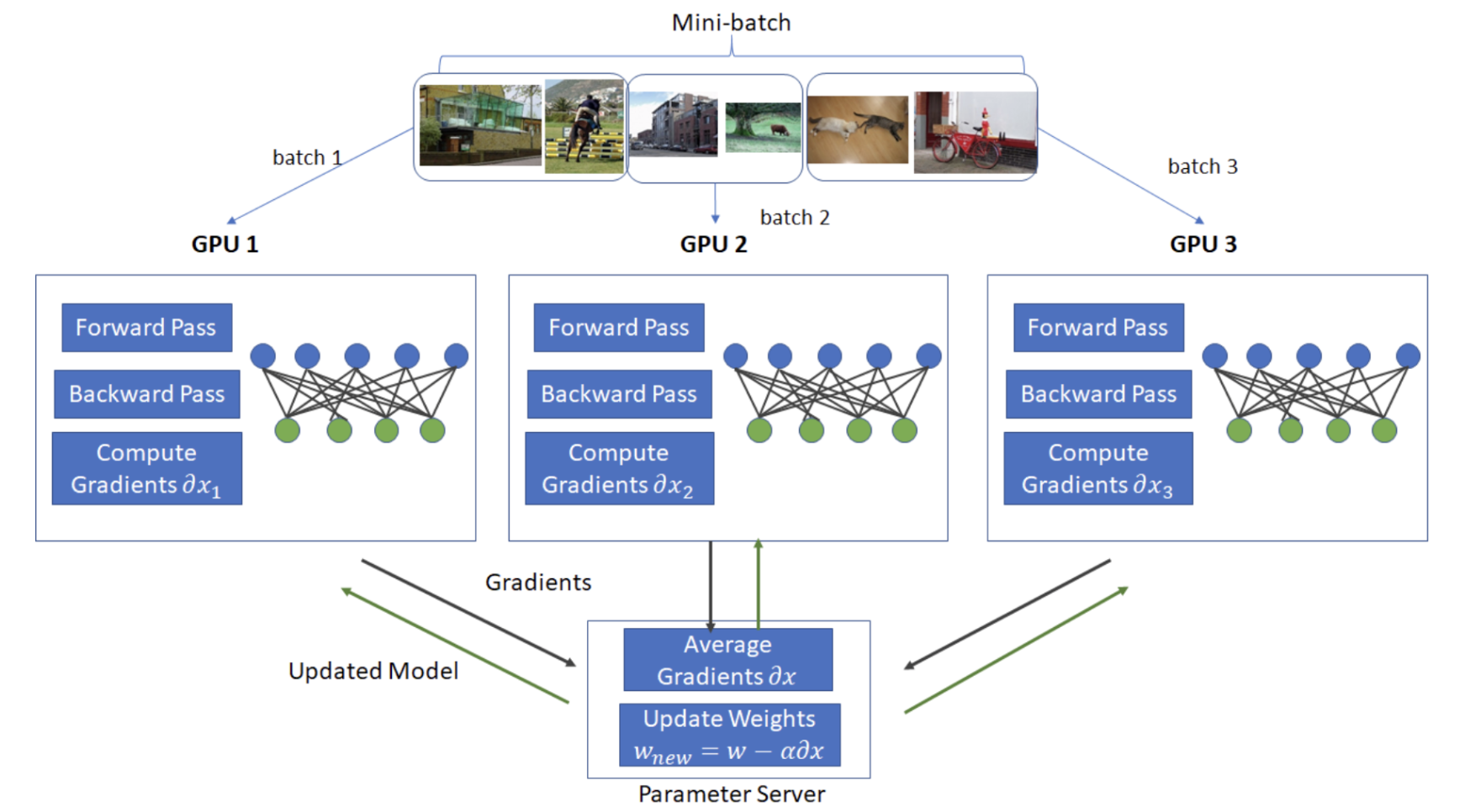}
	}
	}
\subfigure[Model parallelism]{\label{fig:model_paral}
\resizebox{0.99\linewidth}{!}{
	\includegraphics[width=\textwidth]{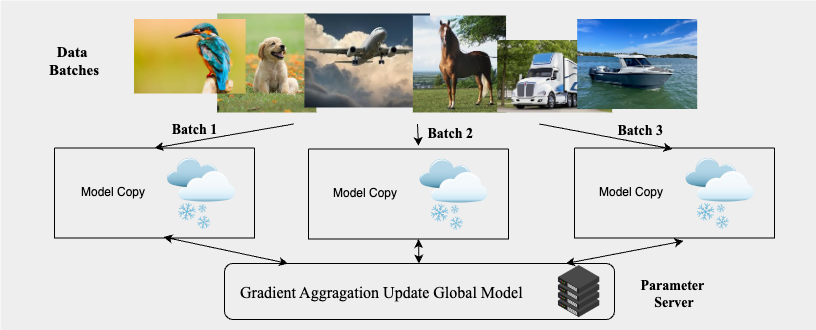}
	}
	}
\caption{Parallel training strategies for large deep neural networks using remote computing nodes.}    \label{fig:parallelism}
\vspace{-0.4cm}
\end{figure}

\subsubsection{Model parallelism}
Model parallelism generally includes two strategies, tenor parallelism and pipeline parallelism. Tensor parallelism divides the model horizontally and assigns each segment of the tensor to a specific GPU. Each GPU independently processes its allocated chunk, allowing for parallel computation across multiple computing nodes. Pipeline parallelism involves distributing each layer (or multiple layers) onto separate computing nodes, either vertically or at the layer-level. The drawback of pipeline parallelism is that it can lead to significantly low computing utilization during the training process. One important technique of Pippy~\cite{pippy2022} automatically splits model layers and converts them into a pipeline across computing nodes. It executes model code concurrently using micro-batches, which helps to alleviate low computing utilization to some extent. Megatron-LM~\cite{DBLP:journals/corr/abs-1909-08053} employs many techniques similar to Pippy for splitting models and mini-batches. Additionally, it introduces activation checkpointing, which stores activations only at the edges of layers. The other activations are recomputed during backpropagation to reduce memory. 

Recently, model parallelism-based federated learning~\cite{DBLP:journals/corr/abs-2004-12088} has been introduced to address data privacy and resource-constrained scenarios. In this approach, deep learning models are split into server-side and client-side components. The server handles the heavy computational tasks, while the client contains a small portion of model shards, performing low-resource computations such as data encoding. This setup ensures that clients have no knowledge of the global model and maintains the privacy of their local data.



\subsection{Blockchain and Benefits}

The above parallelism training methods help distribute the workload and harness the collective processing power of multiple nodes. Commonly, when distributing data or models over networks, encryption should be applied. Local training units encrypt their data and model parameters and send them to the aggregation server. The server performs calculations on the received encrypted data and parameters to obtain the aggregated global model parameters. The server then sends the global model parameters back to the local training nodes, who decrypt the global model parameters locally and proceed with the next round of training. Blockchain is well-suited for situations requiring encrypted data exchange.

Blockchain technology sequentially links blocks, storing transaction records in a chain-like structure and utilizing cryptographic algorithms to ensure the immutability and authenticity of the blocks~\cite{BERDIK2021102397}. Each block consists of two main components: the block header and the block body. The block header contains identifiers for the previous, current, and next blocks, a timestamp, and the Merkle root of transactions forming a Merkle tree within the block body. The block body comprises transactions generated in the blockchain network during a specific period. Transactions are pairwise hashed to generate a hash that serves as the Merkle root, which stored in the block header. Any tampering with transactions alters the Merkle root, ensuring the tamper resistance of block transactions. The block identifier is created by comprehensively calculating the block content and timestamp, ensuring uniqueness and tamper resistance. Blocks are linked through identifiers, ensuring that any change in block content results in a corresponding change in the identifier. Transaction records are broadcasted across the network, enabling nodes to inspect transaction content within each block and ensuring traceability.

Blockchain technologies have also been incorporated into federated learning research recently by utilizing their persistent privacy, security, and decentralized design. Stacey Truex et al.~\cite{10.1145/3378679.3394533} proposed LDP-Fed, which combines local differential privacy (LDP) with federated learning to maintain privacy guarantees during large-scale neural network training. Within the LDP-Fed, the LDP module offers formal differential privacy guarantees for repeatedly collecting model training parameters on private datasets from remote computing nodes during joint training of large-scale neural networks. Laraib Javed et al.~\cite{article1111} utilised blockchain and Local Differential Privacy to construct a secure and reliable data-sharing framework. It creates a trustless environment which utilizes Federated Learning (FL) and Interplanetary File System (IPFS) for decentralized, secure data-driven learning. Feng et al.~\cite{9551794} proposed a BC-FL system for UAVs, where the blockchain is maintained only by entities with high computing and storage capabilities, such as base stations and roadside nodes. This approach utilizes smart contracts to replace the traditional parameter server, enabling transparent and automated model aggregation operations.



\section{Our Model}
\label{sec-model}

\subsection{Identified Challenges}

When training a large language model such as Llama2 70B with half-precision parameters, batch size 32 and 4096 context size, the model with a single batch of data could occupy 150GB of GPU memory. Assuming we have multiple GPU servers with NVidia 4090 GPUs, each with 24GB memory, this training process would require 7 GPU servers to implement pipelined model parallelism, as the model cannot fit into a single GPU card. Moreover, to improve training speed, data parallelism can replicate multiple model parallelism pipelines with different split datasets. Consequently, training a large model could require up to 7 GPUs, which may be not affordable.

Consider a scenario where there is a lack of sufficient computing resources, necessitating the sourcing of additional resources from private computing units. In this situation, it is important to address the following critical security and efficiency issues:

\begin{itemize}

\item Manually model adjustment and maintenance: Adapting a large model for distributed training often necessitates manual changes to the model implementation, such as sharding layers, parameters, and data splitting. This additional workload can burden model development and maintenance, particularly when updating models.

\item Privacy and security issues during the transmission of the model parameters and training data: The model's layers and datasets are transmitted from clients to remote computing nodes. During this transmission over the public network, there is a risk that malicious nodes could modify or intercept the data. This could potentially lead to inferences about the training data distribution or unauthorized replication of the model through data manipulation.

\item Privacy and security issues faced by the global model and back-propagation: Malicious nodes have the capability to perturb local gradients in their local models, potentially misleading the aggregation process and disrupting the convergence of the global model's training.

\item Rewards on faked computing resources: The malicious nodes could deceive the system by generating fake tensor outputs for the next computing nodes and fake gradients for the global server in order to obtain rewards. To prevent fraudulent activity and ensure that rewards are only issued for genuine computations, a robust verification process should be implemented.

\end{itemize}


\subsection{Our Framework}

We proposed a framework to address the aforementioned issues by leveraging blockchain technology to coordinate and secure the processes of LM training and data transmission. Our framework primarily consists of three components: (i) blockchain-based data parallelism; (ii) blockchain-based model parallelism; and (iii) gradient-based malicious node detection.

\subsubsection{Blockchain-based data parallelism}
\label{section:blc_data_paral}

\begin{figure}[h!]
\centering
  \includegraphics[width=\linewidth]{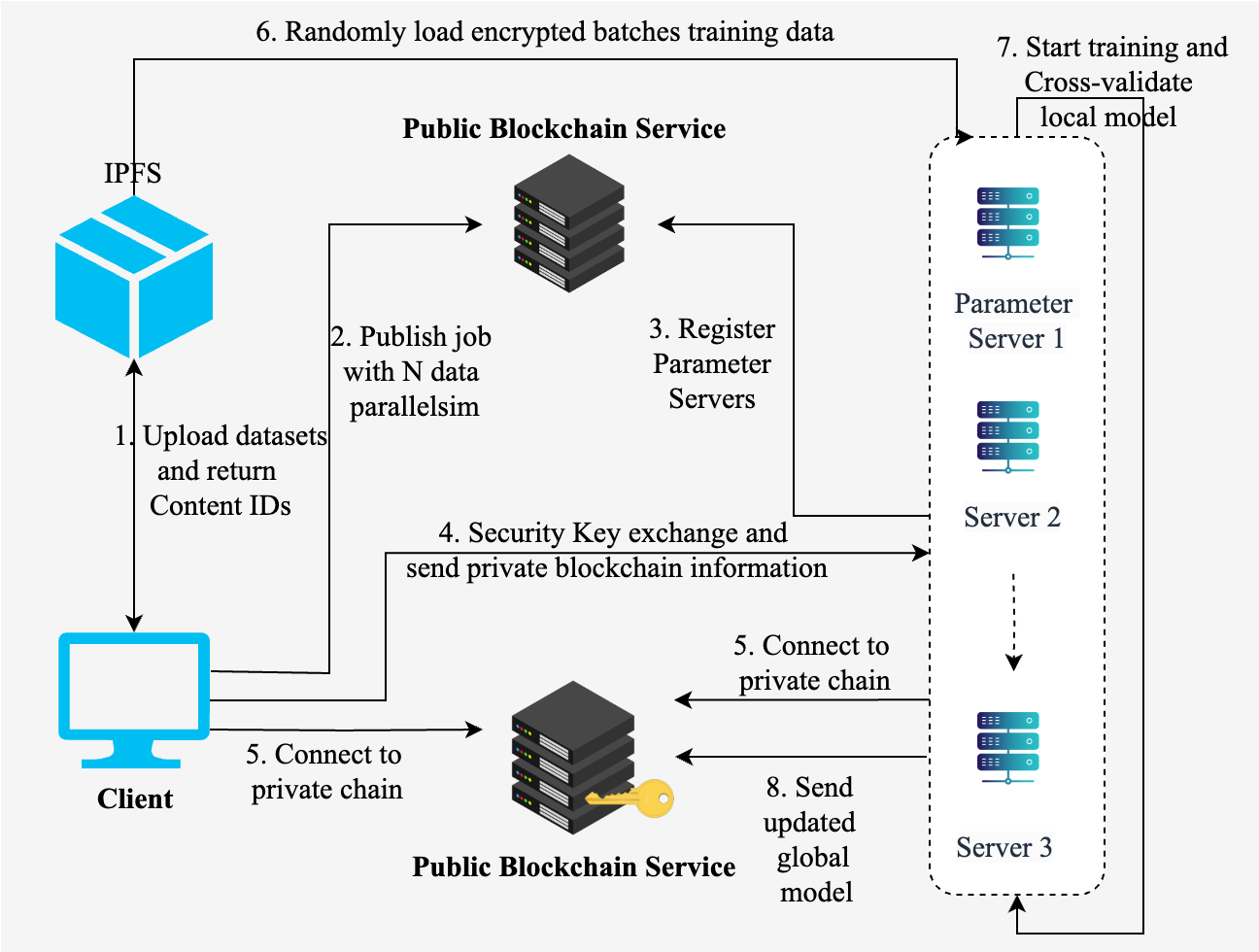}
  \caption{Our blockchain-based data parallelism workflow}
  \label{fig:our_data_framework}
\end{figure}

The blockchain-based data parallelism framework (Figure~\ref{fig:our_data_framework}) consists of independent training pipelines and a validation procedure for the new global model. It includes the following steps: 

\begin{enumerate}

\item Client $C$ initializes the training context by packaging datasets into batches and sending them to an IPFS file server. The IPFS file server returns content IDs (hash codes) for each batch of data to the client.

\item The client publishes a job request with $N$ independent parameter servers for N data pipelines (for more details, please refer to Sec.\ref{section:task_publishing}).

\item Parameter servers register their basic information with the public blockchain service.

\item The client selects N out of the total requests and sends key exchange information. After the key exchange, the client sends encrypted private blockchain information to connect the parameter servers.

\item Once the parameter servers are connected to the private blockchain, they independently initiate their pipeline workflows for pipeline model parallelism training (for more details, please refer to Sec.\ref{section:blc_model_paral}.

\item The parameter servers will load encrypted training data via Content IDs, start local model training, and upload the trained local model information in each epoch to the private chain for cross-validation by other parameter servers by using the test dataset.

\item During the validation procedure, an IDLE parameter server, which has finished its current epoch training and is waiting for other parameter servers, will be assigned as the validating parameter server. The validation server verifies the updated local models published by other parameter servers in the private chain. It creates transactions of the test results to the private chain.

\item After the last local model of the current epoch training has been verified, the validation server aggregates a new global model by averaging the top-$K$ best local models. If the test results are significantly below the mean value, our malicious detection process is triggered. (for more details, please refer to the malicious node detection in Sec.\ref{section:malicious}).

\end{enumerate}

\subsubsection{Blockchain-based model parallelism}
\label{section:blc_model_paral}

\begin{figure}[h!]
\centering
  \includegraphics[width=\linewidth]{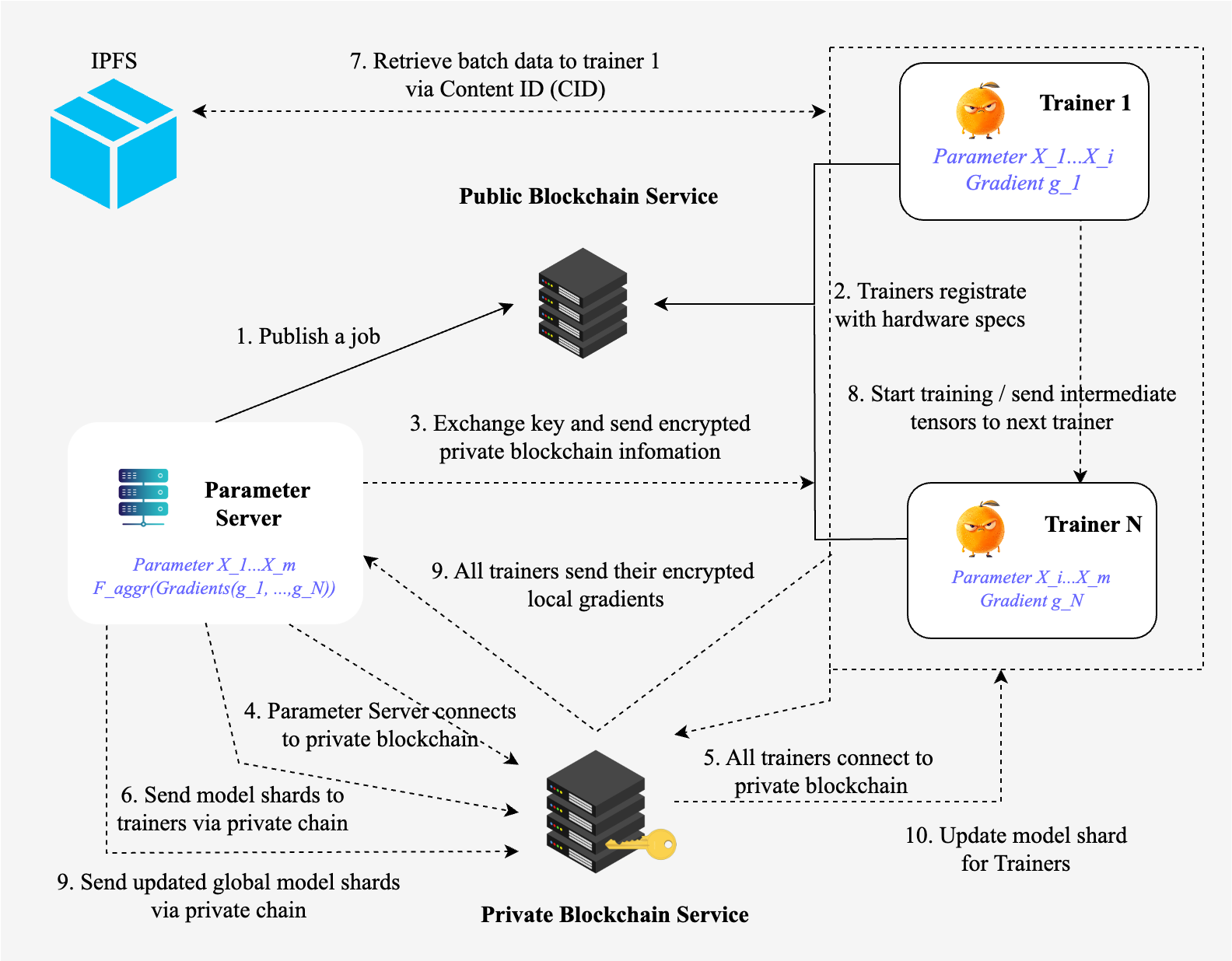}
  \caption{Our blockchain-based model parallelism workflow}
  \label{fig:our_pipeline_framework}
\end{figure}

The blockchain-based model parallelism (Figure~\ref{fig:our_pipeline_framework}) addresses security and malicious detection issues in traditional model parallelism over public networks. It includes the following steps: 

\begin{enumerate}

\item Once the parameter server has been configured by the client's job request, it initiates its pipeline by publishing a computing job request on the public blockchain.

\item Remote computing nodes (trainers) register with their hardware specifications and network details.

\item The parameter server analyzes the global model against the hardware specifications of the candidate list and determines the number of trainers needed. It then selects the top-$K$ trainers whose hardware specifications meet the training job requirements. Then it exchanges keys with the $K$ remote trainers and sends the encrypted private blockchain information to trainers.

\item The parameter server connects to the private blockchain and automatically splits the layers of the global model along with its structural graph according to the specifications of the $K$-selected trainers without needing to modify the original model code. 

\item Once the trainers received the private blockchain information from the parameter server. They will register their specs and information into the private blockchain. 

\item The parameter server uploads the encrypted model shards to the trainers, who then initialize their layers and corresponding parameter weights. 
 
\item The trainer \#1 starts the training loop by randomly loading a new batch from IPFS via CID.

\item The edge layer of each trainer sends the intermediate output tensors to the next trainer in the model's structural graph via the encrypted RPC connection. 

\item Once the epoch is completed, trainers encrypt their local gradients and transmit them to the parameter server through private blockchain transactions. The parameter server then aggregates gradients from remote trainers, computes local gradients using the loss function, updates local model parameters, and subsequently uploads its updated local model to the private blockchain for cross-validation.

\item The parameter server switches to validation mode when there is a pending validation job in the private chain. Otherwise, it proceeds to the next epoch. In the validation model, the parameter server follows Step7 in Sec.\ref{section:blc_data_paral} and publishes the updated global model to a private blockchain. Subsequently, all parameter servers will partition the new global model among their respective trainers for the next epoch.


\end{enumerate}

\subsubsection{Client and Parameter Server Task Publishing}\hfill\\
\label{section:task_publishing}

\noindent\textbf{Client task publishment.}
A client $C$ can initiate a remote ML model training task by publishing an initial task transaction on the public blockchain. The task announcement will include the following information:

\begin{itemize}
\item A timestamp;
\item client UUID
\item A name of this task;
\item A reward budget for the remote computing contributors;
\item A baseline computing capability (e.g., the minimum GPU memory size and the number of CPUs);
\item Determine the number of splits in data-parallel training;
\end{itemize}

Once the task announcement is published on the public blockchain, parameter servers register their basic information in a transaction on the public blockchain service, including a timestamp and their UUID in the chain service.

\smallskip
\noindent\textbf{Parameter server coordinating with remote trainers.}
The second part involves the selected parameter servers gathering remote computing trainers' specifications by sending a training hiring message to the public blockchain, such as:
\begin{itemize}
\item A timestamp;
\item parameter server UUID;
\item A baseline computing capability;
\item A reward budget for the remote computing contributors.
\end{itemize}

The remote computing candidates will evaluate the reward budget and computing requirements. If their hardware meets the baseline criteria, they respond with their hardware specifications. The parameter server then analyzes the global model against these specifications to determine the number of trainers needed, selecting the top-$K$ candidates whose hardware specifications best meet the training job requirements.

\begin{equation}
\mathbb{L} = \bigcup_{l}^{\left \| L \right \|}(\sum_{w}^{\left \| W \right \|}w\times a)
\end{equation}
The above equation lists the memory consumption of each layer of the LM model, where $a$ represents the "bytes per parameter", which is 4 for full precision or 2 for half precision.

After selecting the top-$K$ trainers as the training candidates. The task detail will be published on the private blockchain chain for its remote trainers. The following information will be involved:
\begin{itemize}
\item A timestamp;
\item parameter server UUID
\item A name of this task;
\item A list of shred models of the global model and model structural graph (encrypted);
\item A reward expressed for the computing contributors.
\end{itemize}


\subsection{Security Analysis}
\subsubsection{\textbf{Gradient-based Malicious Node Detection}} \hfill\\
\label{section:malicious}

During distributed training, all trainers send their local gradients to the parameter server for aggregation into the global model. If any local gradients are perturbed or erased, the updated global model could produce undesirable outcomes. In our framework, all transmissions between trainers and the parameter server are secure and encrypted. However, a malicious node could still masquerade as a computing node and execute gradient attacks. Recently, Byzantine attacks have become a focal point of research in the FL domain. These attacks aim to devastate the performance of the global model by manipulating gradient values. The popular methods to defend Byzantine attacks fall into two main categories: (i) malicious node detection and (ii) making the global model more robust to these attacks. In our framework, detecting malicious nodes is more complex than in typical FL problems because remote trainers, acting as malicious nodes, can manipulate many elements during the training steps to alter the global model. Therefore, we designed a two-stage malicious discovery process inspired by the MANDERA method~\cite{DBLP:journals/corr/abs-2110-11736}. 

We assume the process in the $t$-th epoch in trainer M as follows:
\begin{equation}
h_{m}^{t+1} = f_{m}(h_{m}^{t}, w_{m}^{t})
\end{equation}
where $h_{m}^{t+1}$ is the tensor input for the local model layer $t+1$. 
 $h_{m}^{t}$ and $w_{m}^{t}$ are the inputs and parameter weights at the layer $t$. 

During the back-propagation, the parameters $W_m$ of the trainer M calculate the local gradient $g_{m}^{t}$ based on the current layer weight $w_{m}^{t}$, loss function $L_{m}^{t}$ and its local data $h_{m}^{t}$.
\begin{equation}
g_{t}^{m} = \frac{\partial }{\partial w_{m}^{t}} L_{m}^{t}(w_{m}^{t}, h_{m}^{t})
\label{eq:gradient_compute}
\end{equation}

When one epoch training is completed, the parameter server receives gradients from all trainers and updates the global model:
\begin{equation}
W^{t+1} = W^{t} - \gamma_{t} \cdot A(g_{1}^t, g_{2}^t, ..., g_{M}^t)
\label{eq:gradient_aggr}
\end{equation}
where $\gamma_{t}$ is the learning rate.

As Equation~\ref{eq:gradient_compute} shows, the malicious node pretends to be a trainer within the pipeline, which could manipulate the parameter weight $w_m$, the output tensor $h_m$, or the gradient values in Equation~\ref{eq:gradient_aggr}.

To make the training procedure tolerant to malicious nodes and to detect them, we introduced three techniques: (i) a cross-validation method and test dataset for our data parallelism; (ii) top-$N$ local model aggregation to filter out poorly performing local models, which also removes attacked models; and  (iii) malicious node detection at the trainer levels.

The algorithm~\ref{algo:malicious_algo} indicates the parameter server $PS$ enters the validation mode after its training procedure. It will testify the unvalidated local model $M_n$ in the private chain against the test dataset $D$. As Figure~\ref{fig:our_data_framework} shows in Step7, each parameter server (1...$N$) can initiate cross-validation if a new local model is released into the private chain service with validation pending.

\begin{itemize}

    \item  The cross-validation parameter $PS$ server will load the local model into its pipeline, compute against the test dataset $D$, and report the performance results $H$ back to the private chain service. If one or a few results are significantly below the majority, those models will be identified as malicious models $M_{malicious}$ and the malicious node detection procedure will be triggered.

    \item Assuming LM has $||L||$ layers, once the malicious node detection process is triggered, the validation parameter server with the suspicious local model will randomly select $K$ local models plus the suspicious malicious model and project the layers' gradients $G_{li}$ of these models into the lower-dimensional space $R$. Then, we apply $K$-means clustering across the gradients $W_{l}$ of each layer $L_{i}$ to form two groups based on the lower-dimensional space vectors $V_{R}$.

    \item After all local models from parameter servers (1...$N$) have been validated at epoch $i$, the global model aggregation process selects the top-$N$ best-performing local models and averages their weights to form a new global model $M_{global}$. This process filters out local models from any malicious pipelines.

\begin{equation}
L_{malicious} = \underset{L}{argmax}(\frac{Kmeans(G_{i}^{L})}{\sum_{j, i\neq j}^{L}Kmeans(G_{j}^{L})})
\end{equation}

As the above equation indicates, the manipulated gradients should be well-distanced from those of benign nodes. The index of the output of the malicious layer $L_{malicious}$ can then be traced back to the corresponding remote malicious trainers. Finally, the parameter server can remove the malicious trainers from its pipeline.

\end{itemize}

\begin{algorithm}
\caption{Gradient-based Malicious Node Detection}
\small
\begin{algorithmic}[1] 
    \Require Validation Parameter Server $PS$, Test Dataset $D$, and Local Models $M_{n}$ with $||L||$ layers;
    \While{$||M_{n}|| \neq 0$}
        \State $H \gets PS(M_{i}, D)$
    \EndWhile
    \If{$M_{malicious} = argmax(\frac{\left |h_i-\mu\right |}{\sum_{j, j\neq i}^{\left \| H \right \|} \left |h_j - \mu\right |})$}
    \State Trigger malicious node detection process
    \State PS samples models: $M_k + M_{malicious}$  

    \While{$L_{i} \neq ||L||$}
        \While{$M_{i} \neq ||M||$}
            \State Gradient $G_{i}^{L} = \frac{\partial }{\partial w_{i}} L_{i}^{t}(w_{j}^{m}, h_{i}^{m})$
        \EndWhile
        \State Project $R_{i}^{L} \leftarrow  G_{i}^{L}$
    \EndWhile
    \State $\mathbb{L}_{malicious} \gets \underset{L}{argmax}(\frac{Kmeans(R_{L})}{\sum_{j, i\neq j}^{L}Kmeans(R_{L})})$
    \State Map $\mathbb{L}_{malicious}$ to trainers $T_{malicious}$ and Block $T_{malicious}$
    \EndIf

    \State Average aggregate global model: $M_{global} = \mathbb{AGGR}(TopK(M))$
    
\label{algo:malicious_algo}
\end{algorithmic}      
\end{algorithm}

\subsubsection{\textbf{Proof of Training}}\hfill\\
\label{section:pot}

The consensus algorithm of the blockchain guarantees the correct addition of blocks to the blockchain. Its main objective is to establish consensus among the different trainers in the distributed system regarding the system state or transactions. As a distributed system, blockchain utilizes consensus algorithms to tackle issues stemming from network delays, node failures, or malicious activities, ensuring the system's consistency, reliability, and security.


During each training cycle, all remote trainers download the parameters of their sharded models and upload their gradients and local parameters to their parameter server. These local models are then evaluated by the cross-validation parameter server against the test dataset. The validation results and all trainers' outputs are stored in a transaction on the private blockchain in each epoch, checking Sec.\ref{section:blc_data_paral}. 

Therefore, the details of every training procedure are well-recorded by the private blockchain. The entire training process is reproducible, ensuring that the global model is secure and reliable, as the data used, remote trainer activities, and parameter server validation results are all traceable. 


\subsubsection{\textbf{Incentive}}\hfill \\

Once the entire training process is completed and all validation tests are passed, the transactions of all trainers and parameter servers are traceable. The clients can then determine the rewards for the completed computing activities for both the trainers and parameter servers by reviewing the private chain. As a result, this incentive mechanism can significantly encourage clients to behave honestly, given the low likelihood of success and the high risk costs associated with leveraging attacks, which are not worth the potential consequences as analyzed in Sec.\ref{section:malicious} and Sec.\ref{section:pot}.


\section{Experiments}
\label{sec-exp}

To demonstrate the effectiveness of our proposed distributed training method, we conducted a comparative analysis against three baseline techniques on the ResNet50 model with the Cifar-10 dataset. Our experiments aim to achieve three main objectives:

\begin{itemize}
\item We aim to confirm that our training framework, which utilizes a blockchain backbone, can match the performance of single-node training. To achieve this, we need to select an ideal ML model that can fit into a single GPU's memory while also having large enough layers for distributed learning, such as ResNet-50.

\item Federated learning often shows lower accuracy compared to single-node training due to its convergence dependency on the diversity of client data and the parameter averaging process. However, our framework, which utilizes customized data and model parallelism and includes a test dataset for model evaluation, is expected to achieve comparable performance in the assessment.

\item We aim to evaluate the efficiency of our method in learning from distributed training data by examining training loss and convergence speeds. Our framework offers advantages in gradient and model aggregation, ensuring consistent convergence in both training loss and model accuracy. In contrast, federated learning performance is influenced by the number of participating clients and the distribution of their data, resulting in varied convergence curves, especially with fewer clients involved.
\end{itemize}

To verify the above objectives, the experiments are designed with the following configurations:
\begin{itemize}
\item We use ResNet50 and the CIFAR-10 dataset with fixed-split training and test sets across all experiments, including single-node training, FedAvg, and our framework. The same learning rate of 0.1 is applied.
\item To evaluate the quality of model training with multiple computing nodes, we assess FedAvg and our framework with 2, 4, and 8 data parallel trainings. We will present the training loss and accuracy results to compare their performance.
\item We also assess the impact of model sharding on training performance. As the number of computing nodes increases, the layers of ResNet50 will be distributed across more nodes. 
\end{itemize}

\begin{figure}[!h]
  \includegraphics[width=\linewidth]{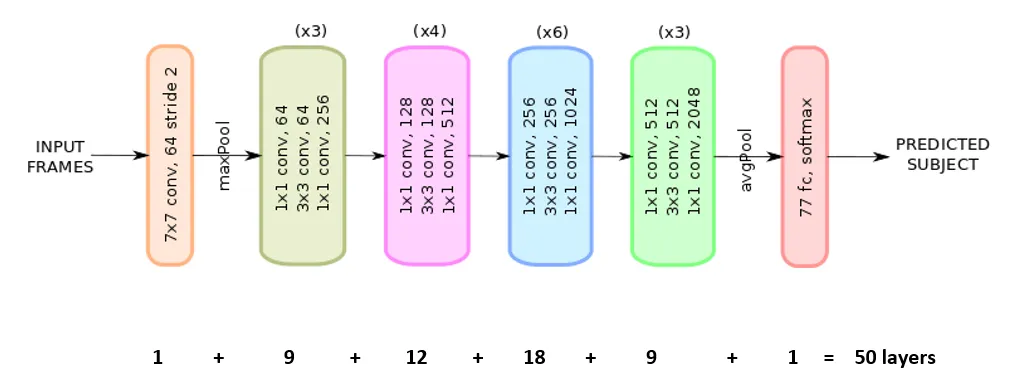}
  \caption{50 layers ResNet architecture}
  \label{fig:resnet50}
\end{figure}


\subsection{Single Computing Nodes and Pipeline Model Parallelism}



We trained ResNet-50 on the CIFAR-10 dataset to achieve performance using the local computer with an Nvidia A6000 GPU. Then, we split the layers of ResNet-50 into 2, 4, and 8 shards, using the PyTorch distributed package to load them onto different GPU servers to test the performance of model parallelism.

\begin{figure}[!h]
\begin{minipage}[h]{0.45\textwidth}
\includegraphics[width=\linewidth]{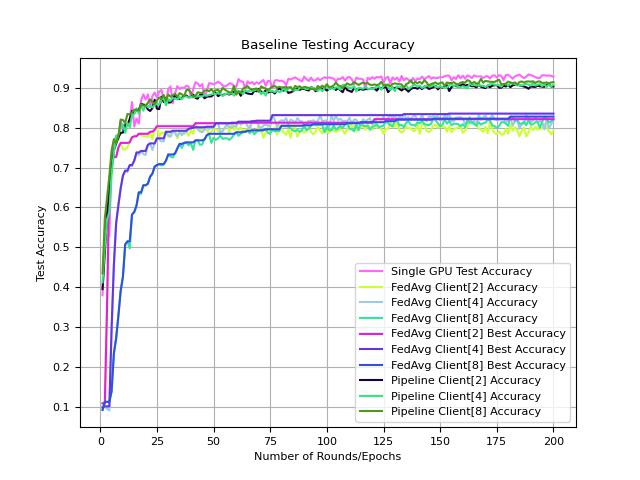}
\caption{Baseline testing accuracy}
\label{fig:baseline_accuracy}
\end{minipage}
\hfill
\begin{minipage}[h]{0.45\textwidth}
\includegraphics[width=\linewidth]{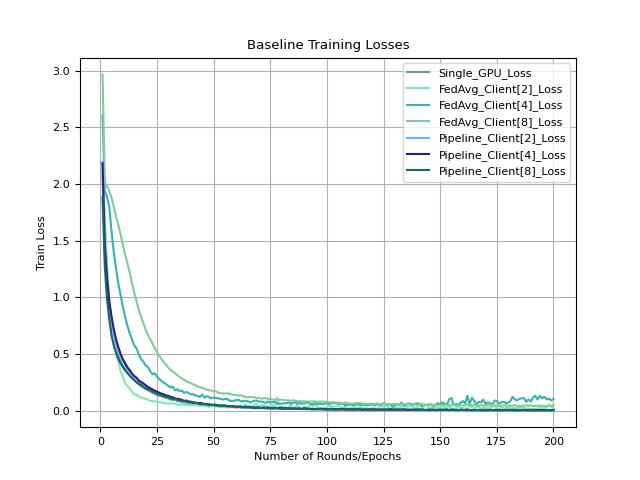}
\caption{Baseline training loss}
\label{fig:baseline_loss}
\end{minipage}
\end{figure}

As Figure~\ref{fig:baseline_accuracy} demonstrates, the pipeline model achieves comparable accuracy performance to single GPU training. In the figures~\ref{fig:baseline_loss}, it converges at a similar speed to the single GPU model with similar accuracies when comparing the black line (Pipeline\_Client[4]\_Loss) and the grey line (Single\_GPU\_Loss). 

Note: In "Pipeline\_Client[x]", "x" indicates the number of remote computing units, which can be "2", "4", or "8" in our experiments. This notation applies similarly in the following experiments.

\begin{figure}[h]
\begin{minipage}[b]{0.45\textwidth}
\includegraphics[width=\linewidth]{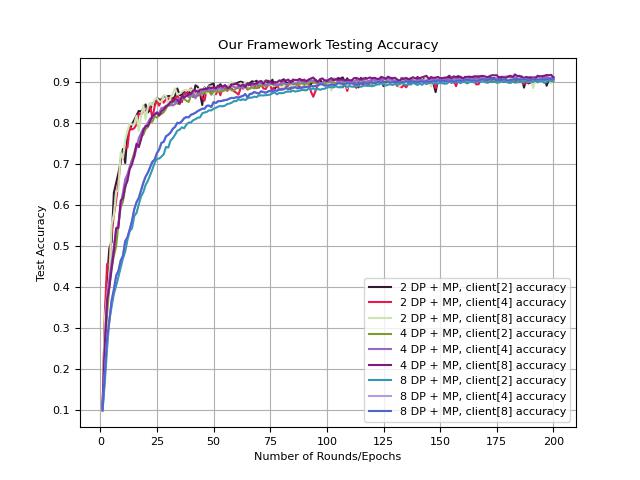}
\caption{Our framework's testing accuracy}
\label{fig:our_accuracy}
\end{minipage}
\hfill
\begin{minipage}[b]{0.45\textwidth}
\includegraphics[width=\linewidth]{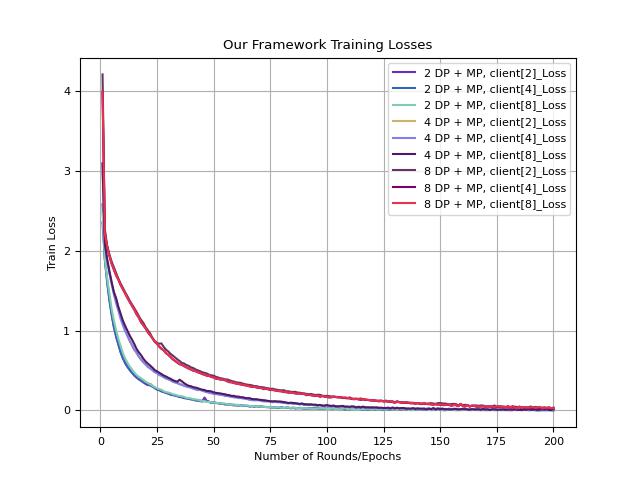}
\caption{Our framework's training loss}
\label{fig:our_loss}
\end{minipage}
\end{figure}

\subsection{Federated Learning}

The FedAvg training performance is evaluated by using different numbers (2, 4, 8) of remote computing nodes (clients). Figure~\ref{fig:baseline_accuracy} shows that the accuracy of the FL models on CIFAR-10 is significantly lower than that of the pipeline model and the single GPU model. Both the best and real-time accuracies of FedAvg only reach 83\%. Additionally, as the number of remote computing nodes increases, the training losses remain higher and converge more slowly compared to the other two baselines.



\subsection{Our Solution}
Figure~\ref{fig:our_accuracy} and Figure~\ref{fig:our_loss} demonstrate our framework performance. Since our framework includes data parallelism (DP) and pipeline model parallelism (MP) for large model training, the report schema has two parts: (i) [m] DP and (ii) MP client[$N$]. The first part, similar to FedAvg, divides the dataset into M pieces with independent model training and aggregates the local models into a global model after each epoch. The second part, MP client[$N$], represents the pipeline model parallelism training where the entire model is split into $N$ shards to ease memory requirements. For example, "2 DP + MP, client[2]" means the training set is divided into two with independent models, and each model is split into two shards across two computing nodes.

Regarding the model converging speed, Figure~\ref{fig:our_loss} shows a smooth curve with decreasing losses. Due to the aggregation process in the data parallelism method, it also experiences similar problems as FedAvgwith convergence speed decreasing as more [M] DP is involved. However, our top-$K$ aggregation method results in smoother convergence loss compared to FedAvg. Compared to the figures~\ref{fig:baseline_loss}, FedAvg losses spike after epoch 150. Regarding the determination of the $K$ value, $K=M/2$. 2 DP is $K=1$; 4 DP is $K=2$; 8 DP is $K=4$.

Our model achieves a similar performance of 90\% accuracy to that of a single computing node, as shown in Figure~\ref{fig:our_accuracy}. However, similar to the issues faced with FedAvg DP, a larger number of DP negatively impacts convergence performance. The blue line (8 DP + MP, client[8]) shows lower accuracy than setups with fewer DPs, reaching 90\% accuracy after 100 epochs, whereas others achieve this after 75 epochs.

\subsection{Malicious Nodes Detection}

In Sec.\ref{section:malicious}, we present our detection methods for trainers under Byzantine attacks. In this section, we illustrate the behavior for various types of attacks, including Gaussian attacks, Zero gradient attacks, and Mean shift attacks. 


\begin{figure*}
\centering
\subfigure[Zero gradient attack]{\label{fig:zero}
\resizebox{0.3\linewidth}{!}{
	\includegraphics[width=\textwidth]{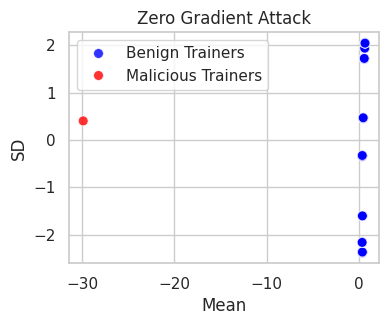}
	}
	}
\subfigure[Mean shift attack]{\label{fig:mean}
\resizebox{0.3\linewidth}{!}{
	\includegraphics[width=\textwidth]{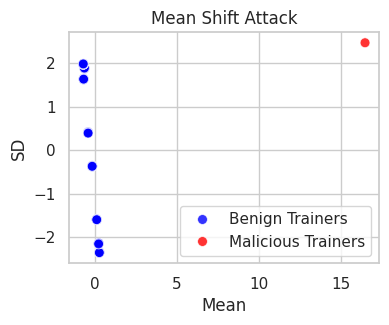}
	}
	}
\subfigure[Gaussian attack]{\label{fig:gaussian}
\resizebox{0.3\linewidth}{!}{
	\includegraphics[width=\textwidth]{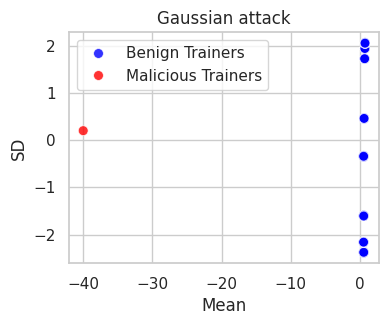}
	}
	}
\caption{The mean rankings and standard deviations for benign and malicious trainers after one epoch has completed.}
\label{fig:attack}
\end{figure*}

\subsubsection{Zero gradient attack}

A zero gradient attack aims to make the aggregated message to be zero, such as $G_{i}^{L} = \frac{\partial }{\partial w_{i}} L_{i}^{t}(w_{j}^{m}, h_{i}^{m}) = 0$. Therefore, the next neighbour layer weights will stop the learning process. Figure~\ref{fig:zero} shows an illustrative motivation to our method. It demonstrates that our method clearly separates malicious nodes from benign nodes under zero gradient attacks. 

\subsubsection{Mean shift attack} 

A mean shift attack~\cite{10.5555/3454287.3455062} injects minor noise to poison the trainers' gradients to disrupt the learning updates of parameters. Because this attack is not well-identified, it is challenging to distinguish from true gradient distributions. However, such attacks rarely occur in our framework. The trainer is unaware of the actual gradient distribution and must contribute to the learning process before launching an attack which makes the cost for attackers higher than simply acting as benign nodes. Figure~\ref{fig:mean} demonstrates that our method clearly separates malicious nodes from benign nodes under mean shift attacks. 

\subsubsection{Gaussian attack} 

In a Gaussian attack, the attacker generates random malicious gradient values with a Gaussian distribution covariance matrix $\Sigma_m$. Similarly to the mean shift attack, the attackers need to analyse the gradient distribution ($\mu_b, \Sigma_b$) before introducing $\Sigma_m$ noise. The small $\Sigma_m$ noise could not affect our framework performance because its attack may located in the gradient distribution  ($\Sigma_m \subset \Sigma_b, where\ \Sigma_b = \frac{1}{\left \| b \right \|}\sum_{i}^{b}s_{i}^{2}$). If the large $\Sigma_m$, it would be identified easily. Moreover, our framework will only aggregate the Top-K local models and the minor Gaussian attacks can be filtered out easily. Overall, It is a very costly attack method for our framework. During this experiment, we follow the configure of MANDERA~\cite{DBLP:journals/corr/abs-2110-11736} to set $\Sigma = 30I$ for the Gaussian attack. Figure~\ref{fig:gaussian} shows that our method effectively separates Gaussian attacks from benign nodes.

\section{Conclusion}

The limitation of computing resources has been a significant barrier to the rapid development of large models (LM). Existing techniques have not fully addressed this issue. Blockchain-based federated learning has partially solved data privacy and computing limitations across distributed environments, but it still lacks a comprehensive solution for training large models in a distributed manner.
In this paper, we propose a trustworthy distributed machine learning (TDML) framework to tackle the challenges of LM training over a public (distributed) network.  We integrate blockchain techniques with parallel training methods to ensure secure, trustworthy, and traceable distributed training. Our malicious discovery method and cross-validation mechanism efficiently identify suspicious attacking nodes while verifying the work of computing nodes for the reward system. Further, our framework allows for the contribution of spare computing resources and the earning of rewards. Experimental results demonstrate that our aggregation method improves model performance by over 10\% in accuracy compared to FedAvg and matches the baseline performance of a single computing node.




\bibliography{bib}
\bibliographystyle{splncs04}

\end{document}